\documentstyle[12pt]{article}
\def\beq{\begin{equation}}
\def\eeq{\end{equation}}
\def\ea{\begin{eqnarray}}
\def\beaa{\begin{eqnarray*}}
\def\eea{\end{eqnarray}}
\def\eeaa{\end{eqnarray*}}
\def\bq{\begin{quote}}
\def\eq{\end{quote}}
\def\gappeq{\mathrel{\rlap {\raise.5ex\hbox{$>$}}
{\lower.5ex\hbox{$\sim$}}}}

\def\lappeq{\mathrel{\rlap{\raise.5ex\hbox{$<$}}
{\lower.5ex\hbox{$\sim$}}}}

\begin{document}
\pagestyle{empty}
\vspace*{5mm}

\begin{center}

{\bf ON THE INTERPRETATION OF THE REDSHIFT IN\\ A STATIC GRAVITATIONAL FIELD}\\
\vspace*{0.8cm}
{\bf L.B. OKUN} and {\bf K.G. SELIVANOV}\\
\vspace{0.2cm}
ITEP, Moscow, 117218, Russia \\
{\tt e-mail: okun@heron.itep.ru, selivano@heron.itep.ru} \\
\vspace{0.2cm}
and\\
\vspace{0.2cm}
{\bf V.L. TELEGDI}\\
EP Division, CERN,
CH - 1211 Geneva 23\\
{\tt e-mail: valentine.telegdi@cern.ch}\\
\vspace*{1.8cm}
{\bf ABSTRACT} \\
\end{center}
\vspace*{5mm}

The classical phenomenon of the redshift of light in a static
gravitational
potential, usually called the gravitational redshift,
is described in the literature essentially in two ways: on
the one hand the phenomenon is explained  through the
behaviour of clocks which run the faster the higher they are located
in the potential, whereas the energy and frequency of the
propagating photon do not change with
height. The light thus appears to be redshifted relative to the
frequency of the
clock. On the other hand the phenomenon is alternatively discussed
(even in some
authoritative texts) in terms of an energy loss of a photon as it
overcomes the
gravitational attraction of the massive body.
This second approach operates with notions
such as the ``gravitational mass" or the ``potential energy" of a photon
and we
assert that it is misleading. We do not claim
to present any original ideas or to give a comprehensive review of the
subject, our goal being  essentially a pedagogical one.

\vfill\eject
%\pagestyle{empty}
%\clearpage\mbox{}\clearpage

\setcounter{page}{1}
\pagestyle{plain}

\section{Introduction}

There are two kinds of photon redshift known in the literature, a
gravitational and a cosmological one. Though in the General Relativity
framework the two shifts can be described very similarly, in the literature
they are usually discussed separately.
The cosmological redshift is that of light from distant galaxies
which recede. It is generally referred to as the Hubble redshift.
For the farthest observed galaxies it is quite large,
$\Delta \lambda/\lambda \approx 5$.
The gravitational redshift arises
when light moves away from a static massive object, e.g. the earth or the
sun. Its observed magnitudes are generally small. This paper is devoted
exclusively to this type of redshift.

The gravitational redshift is a classical effect of Einstein's General
Relativity (GR), one predicted by him \cite{aaa} well before that
theory was created \cite{bb} (for the historical background, see e.g.,
\cite{Pais}). Phenomenologically one can
simply affirm that the frequency of light emitted by two identical
atoms is smaller for the atom which sits deeper in the gravitational
potential. A number of ingenious experiments  have been performed
\cite{cc}--\cite{hh} to measure various manifestations of this effect.
They are discussed in a number of excellent reviews \cite{jj} whose main
goal is to contrast the predictions of GR with
those of various non-standard theories of gravity. Explanations of
the gravitational redshift {\it per se} within the standard framework are
however not critically discussed in these reviews.

Most treatises on GR \cite{nn}, \cite{qq} follow the definitive reasoning of
Einstein  \cite{bb} according to which
the gravitational redshift is explained in terms of universal property
of standard clocks (atoms, nuclei). The proper time interval between
events of
emission of two photons as measured by the standard clock at the point of
emission is different from the  proper time interval between events of
absorption of those photons as measured by identical standard clock at the
point of absorption (in this way it was first formulated in \cite{Weil}).

In the static gravitational potential the picture
simplifies because
there is a distinguished time -- the one on which metric is independent.
This time can be chosen as a universal (world) one. Under this choice
the energy difference between two
atomic levels increases with the distance of the atom from the earth
while the
energy of the propagating
photon does not change.  (In what follows we speak of the earth, but
it could be  any other massive body.)
Thus what is called the redshift of the photon is actually a
blueshift of the atom. As for the proper times at different points,
they are
related to the universal time via a multiplier which depends on
gravitational potential and hence has different values at different points
(see section 4).

Actually, most modern textbooks and monographs  \cite{modern}
derive the redshift by using sophisticated general relativity
calculations, e.g. using orthonormal bases (a sequence of proper reference
frames) to define photon energy and parallel transporting the photon's
4-momentum along its world-line. Sometimes this description is loosely
phrased as a degradation of the photon's energy as it climbs  out of
gravitational potential well. (Some other classical textbooks also
use this loose phrasing \cite{19}.) However, the non-experts
should be warned that the mathematics underlying this description is
radically
different from the heuristic (and wrong) arguments presented in many
elementary texts, e.g. \cite{uu}. These authors claim to deduce
the ``work against gravity'' viewpoint by pretending that the photon is
like a normal, low-velocity, massive particle and thus has a ``photon mass''
and ``photon potential energy''. Such derivations are incorrect and should
be avoided. They are in fact avoided in exceptional popular texts,
e.g. \cite{yy}.

\section{Experiments}

The first laboratory measurement of the gravitational redshift was
performed at
Harvard in 1960 by Robert Pound and Glenn Rebka \cite{cc}, \cite{dd}
(with 10\%
accuracy) and in 1964 by Pound and Snider \cite{ee} (with 1\% accuracy). The
photons moved in a 22.5-meter tall tower.
The source and the absorber of the photons ($\gamma$-rays of 14.4 keV energy)
were $^{57}$Fe nuclei. The experiment exploited the M\"ossbauer effect which
makes the photon lines in a crystal extremely monochromatic. The redshift
was compensated through the Doppler effect, i.e., by slowly moving the
absorber
and thereby restoring the resonant absorption. The shift measured in this
way \cite{dd} was $\Delta\omega / \omega \approx 10^{-15}$.

As to the interpretation of the result, there is some ambiguity in the
papers by
Pound et al. Although they mention the clock interpretation by referring to
Einstein's original papers, the absolute reddening of the photon is also
implied as can be seen from the
title of Pound's talk in Moscow \cite{dd} ``On the weight of photons". From
the title of the paper by Pound and Snider \cite{ee}, ``Effect of gravity on
nuclear resonance" one might infer that they did not want to commit
themselves to any interpretation.

By contrast, the majority of the reviews of gravitational experiments
\cite{jj} quote the Harvard result as a test of the behaviour of clocks.
In fact the result must be interpreted as a relative shift of
the photon frequency with respect to the nuclear one since the experiment
does
not measure these frequencies independently.

An experiment measuring the relative shift of a photon
(radio-wave) frequency with respect to an atomic one was also performed
with a rocket flying up to 10,000 km above the earth and landing
in the ocean [7].

Alongside the experiments [3]--[7] special measurements of the
dependence of the atomic clock rate on the altitude were done
directly by using airplanes [8], [9] (see also reviews [10]). In
these experiments a clock which had spent many hours at high altitude was
brought back to the laboratory and compared with its ``earthly twin".
The latter, once corrected for various background effects, lagged
behind by $\Delta T = (gh/c^2)T$, where $T$ is the duration of the flight
at height $h$, $g$ the gravitational acceleration, and $c$  the speed of
light.

One of these background effects is the famous
``twin paradox" of Special Relativity, which stems from the fact that moving
clocks run slower than clocks at rest. It is easy to derive a general formula
which includes both the gravitational potential $\phi$ and
the velocity $u$ (see, e.g., the book by C. M\"{o}ller \cite{nn}):  \beq
d\tau = dt \left[ 1+2\phi /c^2 - u^2/c^2\right]^{1/2}~~, \label{one} \eeq
where $\tau$ is the proper ``physical" time of the clock, while $t$ is the
so-called world time, which can be introduced in the case of a static
potential and which is sometimes called laboratory time.

In his lectures on gravity \cite{qq}, Richard Feynman stresses the
differences between the effects due to $u$ and to $\phi$. He concludes
that the centre of the earth must be by a day or two younger than
its surface.

\section{Theory before General Relativity: ``elevators''}

Since most of the conclusive experiments on the gravitational redshift were
earthbound, we shall throughout use that frame in which the earth is at rest
(neglecting its rotation).

As is well known, a potential is defined up to a constant. When considering
the gravitational potential $\phi(r)$ at an arbitrary distance $r$ from the
earth's centre, it is convenient to set $\phi(\infty) = 0$; then $\phi$ is
negative everywhere.

Near the earth's surface (at $h = r - R \ll R$) it is legitimate to
approximate $\phi$ linearly:
\beq
\delta\phi(h) = \phi(R+h) -\phi(R) = gh \;\; ,
\label{new}
\eeq
where $g$ is the usual gravitational acceleration. Note that $\delta\phi(h)$
is positive for $h \neq 0$.
We shall discuss the redshift only to the first order in the parameter
$gh/c^2$.

(The linear approximation Eq. (\ref{new}) is valid for the description of
experiments \cite{ggg, hh}. It is obvious, however, that for the high-flying
rockets \cite{ff} ($h\approx 10^4$ km) it is not adequate and the newtonian
potential should be used, but this is not essential for the dilemma ``clocks
versus photons'' which is the subject of this paper.)

Einstein's first papers \cite{aaa} on the gravitational redshift
contain many of the basic ideas on the subject which were incorporated
(sometimes without proper critical analysis) into numerous subsequent texts.
He considered the Doppler effect in the freely falling frame and
found the increase of the frequency of an atom (clock) with increasing height
(potential).  The
cornerstone of his considerations was the local equivalence between
the behaviour of
a physical system in a gravitational field and in a properly accelerated
reference frame.

For the potential (\ref{new}) it is particularly convenient to appeal to
Einstein's freely falling  reference frame (``elevator"). In such an
elevator an
observer cannot detect any manifestation of gravity by any {\it strictly
local}
experiment (equivalence principle). Operationally, ``strictly local"
means that the device used is sufficiently small not to be sensitive
to curvature effects.

Consider from such an elevator falling with the acceleration $g$
a photon of frequency $\omega$ which is emitted upwards by an atom at
rest on the surface of the earth and which is expected to be absorbed by an
identical atom fixed at height $h$.
The frequency of light is not affected by any gravitational field in a freely
falling elevator: it keeps the frequency with which it was emitted.
Assume that at the moment of emission $(t=0)$
the elevator had zero velocity.
 At the time $t = h/c$, when the photon reaches the absorbing atom, the
latter will have velocity $ v = g h/c$  directed
upwards in the elevator frame. As a result the frequency of the
photon, as seen by the absorbing atom, will be shifted by the linear Doppler
effect by $v/c=gh/c^2$ towards the red, that is
\beq
\frac{\Delta\omega}{\omega} = -\frac{gh}{c^2} ~~.
\label{two}
\eeq

(Minute corrections of higher order in $gh/c^2$ to the ``elevator formulas"
are lucidly discussed on the basis of a metric approach in ref. \cite{17a}.)
 Consider now another situation, when the upper atom (absorber) moves in the
laboratory frame with a velocity $v = gh/c$ downwards. Then
in the elevator frame it will have zero velocity at the moment of absorption
and hence it will be able to absorb the photon
resonantly in complete agreement with
experiments \cite{cc, dd}. Obviously, in the elevator frame there is no room
for the interpretation of the redshift in terms of a photon
losing its energy
as it climbs out of the gravitational well.

\section{Theory in the framework of General Relativity: metric}

Up to now we used only Special Relativity and newtonian gravity.
As is well known, a consistent relativistic description of classical
gravity  is given in the framework of GR with its curved space-time metric.
One introduces a metric tensor, $g_{ik}(x), i$, $k=0,1,2,3$, which is,
in general, coordinate
dependent and  transforms by definition under a change of coordinates in such
a way that the interval $ds$ between two events with coordinates $x^i$ and
$x^i +dx^i$,
\beq
ds^2 = g_{ik}(x)dx^i dx^k
\label{2}
\eeq
is invariant. Setting $dx^1 =dx^2 =dx^3 =0$,
one obtains the relation between the proper time interval $d\tau
=ds/c$ and the world time interval $dt =dx^0/c$ for an
observer at rest

\beq
d\tau =\sqrt{g_{00}}dt ~~.
\label{3}
\eeq
For a static case, Eq. (\ref{3}) integrates to
\beq
\tau =\sqrt{g_{00}}t ~~,
\label{4}
\eeq
where $g_{00}$ is a function of \mbox{\boldmath$x$}
in the general case while in the case
of Eq.(\ref{new}) it is a function of $x^3=z=h$.

The time $\tau$ is displayed by a standard clock and
can also be viewed on as a
time coordinate  in the so-called comoving locally inertial frame, i.e. such
locally inertial frame which at a given instant has zero velocity with
respect to the laboratory frame. If one has a set of standard clocks at
different points, then their proper times $\tau$ are differently related to
the world (laboratory) time $t$, due to \mbox{\boldmath$x$}-dependence of
$g_{00}$ (see Eq. (\ref{4})).  This explains the airplane
experiments [8], [9].  Let us recall that sometimes the world time is called
 laboratory time. The former term reflects the fact that it is the same for
the whole world, the latter signifies that it can be set with standard clocks
in the laboratory. Many authors refer to $t$ as the coordinate time.

A weak gravitational field can be
described by a gravitational potential $\phi$, and $g_{00}$ is related to the
gravitational potential:
\beq
g_{00} = 1+2\phi/c^2 ~~.
\label{5}
\eeq
We shall explain the meaning of this relation a bit later (see Eqs.
(\ref{6})--(\ref{8})).
According to Eqs. (\ref{4}), (\ref{5}) a clock runs the slower
in the laboratory the deeper it
sits in the gravitational potential.

Analogous to Eq. (\ref{3}) is a relation between the rest energy
$E_0$ of a body
in the laboratory frame and in the comoving locally inertial frame,
\beq
E^{lab}_0 =E^{loc}_0 \sqrt{g_{00}}
\label{6}
\eeq
(notice that $E^{lab}_0 dt =E^{loc}_0 d\tau$; this is because the energy $E$
is the zero-th component of a covariant 4-vector, while $dt$ is the zero-th
component of a contravariant 4-vector).

The rest energy in the locally inertial frame is the same as in the special
relativity (see e.g. \cite{zz} and the book by E. Taylor and J.A. Wheeler
\cite{uu}, p.246).
\beq
E^{loc}_0 = mc^2 \;\; ,
\label{7}
\eeq
while the rest energy in the laboratory system $E^{lab}_0$ includes also
the potential energy of the body in the gravitational field. This is in
accordance with the main principle of the general relativity: all effects of
gravity arise only via the metric tensor. Eq. (\ref{5}) for the $g_{00}$
component of the metric tensor in a weak gravitational field can be
considered as a consequence of Eqs. (\ref{6}), (\ref{7}) and of the relation
\beq
E^{lab}_0 = mc^2 +m\phi ~~,
\label{8}
\eeq
which generalizes the notion of the rest energy of a free particle to that of
a particle of mass $m$ in a gravitational weak potential.

Now we are in a position to explain the redshift in the laboratory frame.
According to Eq. (\ref{6}) or Eq. (\ref{8}) the energy difference
$\varepsilon_{lab}$ of atomic or nuclear levels in that frame
depends on the location of the atom. The deeper atom sits in the
gravitational potential the smaller is $\varepsilon_{lab}$.
For an  absorber atom
which is located at height $h$ above an identical atom which emits
the photon, the relative change in the energy difference is $gh/c^2$,
\beq
\frac{\Delta \varepsilon^{lab}}{\varepsilon^{lab}} = \frac{gh}{c^2} ~~.
\label{9}
\eeq
(We use in Eq. (\ref{9}) the linear approximation of Eq. (\ref{new}).)
One can say that the energy levels of the absorber atoms  are shifted towards
the blue in the laboratory frame. Eq. (\ref{9}) is, of
course, nothing but a way to describe the difference in
the rates of atomic
clocks located at a height $h$ one above the other. On the other hand,
the
energy (frequency)
of the photon is conserved as it propagates in a static gravitational
field. This can, for example, be seen from the
wave equation of electromagnetic field in the presence of a static
gravitational potential
or from the equations of motion of a massless (or massive)
particle in a static metric.
Clearly, in the laboratory system there is no room for the interpretation
in which the photon loses its energy when working against the gravitational
field.

Finally, one can discuss the experiment using a sequence of
locally inertial frames which are comoving with the laboratory clocks (atoms)
at the instants when the photon passes them. As we explained above, in such
systems the standard clocks run with the same rates, the rest energy of the
atom is equal to its mass, times $c^2$,
Eq. (\ref{7}), and the energy levels of the atom are
the same as at infinity. On the other hand, the energy of the photon in
the laboratory frame $E_{\gamma}^{lab}=\hbar \omega^{lab}$
and in the comoving locally inertial frame
$E_{\gamma}^{loc}$ are related as
\beq
E_{\gamma}^{lab} =E_{\gamma}^{loc} \sqrt{g_{00}} ~~.
\label{15}
\eeq
Eq.(\ref{15}) follows from Eq.(\ref{6})
by noticing that the photon can be absorbed
by a massive body and by considering the increase of the rest energy
of that body.
Thus, since $E_{\gamma}^{lab}$ is conserved, $E_{\gamma}^{loc}$
decreases with height:
\beq
\frac{\omega^{loc}(h) -\omega^{loc}(0)}{\omega^{loc}(0)} =
\frac{E_{\gamma}^{loc}(h) -E_{\gamma}^{loc}(0)}
{E_{\gamma}^{loc}(0)} =-\frac{gh}{c^2}
\label{16}
\eeq
and this is the observed redshift of the photon. But $E^{loc}$
decreases not because the photon works against the gravitational field. The
gravitational field is absent in any locally inertial frame.
$E_{\gamma}^{loc}$
changes because one passes from one of the  locally inertial frames to the
other: the one comoving with the laboratory at the moment of emission, the
other -- at the moment of absorption.

\section{Pseudoderivation and misinterpretation of gravitational redshift}

The simplest (albeit wrong) explanation of the redshift is based on ascribing
to the photon both an inertial and a gravitational`` mass"
$m_{\gamma} = E_{\gamma}/c^2$. Thereby
a photon is attracted to the earth  with a force $g m_{\gamma}$, while
the fractional decrease of its energy (frequency) at height $h$ is
\beq
\frac{\Delta E_{\gamma}}{E_{\gamma}} = {\Delta \omega\over\omega}
=  -g m_{\gamma}h / m_{\gamma}c^2 =
-g~h/c^2~.
\label{nine}
\eeq
Note that (up to a sign)  this is exactly the formula for the  blueshift
of an atomic level. That should not surprise. The atom and the photon are
treated
here on the same footing, i.e.  both
non-relativistically! This is of course inappropriate for the photon.
If the explanation in terms of gravitational attraction of the photon to the
earth were also correct, then one would be forced to expect a doubling of the
redshift (the sum of the effects on the clock and on the photon) in the
Pound-type experiments.

Some readers may invoke Einstein's authority to contradict what was said
above.
In his 1911 paper \cite{aaa},
he advanced the idea that energy is not only a source of inertia, but also
a source of gravity. Loosely speaking, he used the heuristic argument
``whenever
there is mass, there is also energy and vice versa". As he realized
later, this
``vice versa" was not as correct as the direct statement (a photon has energy
while its mass is zero). By applying this energy-mass argument, he calculated
the
energy loss of a photon moving upwards   in the potential of the earth as
discussed above. With the same heuristic principle he also derived an
expression for the deflection of light by the sun which however
underestimated the deflection angle by a factor of 2. Subsequently, in the
framework of GR, Einstein recovered this factor \cite{bb}. The correct
formula was confirmed by observation.

\section{The wavelength measurement}

Up to now we have discussed the gravitational redshift in terms of
the photon frequency and clocks.
 Let us now consider the same phenomenon in terms of the photon wavelength
and gratings. To do this we consider two identical gratings at
different heights, inclined with respect to the $z$-axis along which the
light
propagates between them. We do not go into the details of this gedanken
experiment.
The $z$-projections of grating
spacing is used only as a standard of length.  The lower grating serves as
monochromator, i.e.  as the light source.  The wavelength of the photon
${\lambda}^{lab}(z)$ corresponds to its frequency, while the spacings of the
grating in vertical $( z )$ direction $l^{lab}(z)$ correspond to the rates of
the clocks.

For the sake of simplicity, one may consider a very small incidence angle on
the gratings, i.e. the grazing incidence of the light. In that case, the
vertical projection of the spacing is practically the spacing itself. (Recall
that, for the grazing incidence, the spacing $l^{lab}$ must be of the same
order as the wavelength ${\lambda}^{lab}$.)

While the photon energy $E^{lab}$ is conserved in a static
gravitational field the photon momentum $p^{lab}$ is not.
The relation between these quantities is given by the condition
that the photon remains massless which in a gravitational field reads
\beq
g^{ij}p_{i}p_{j}=0 \;
\label{15'}
\eeq
where the $g^{ij}, i,j=0, \ldots ,3$ are the contravariant components
of the metric tensor,
$p_{j}$ are the components of the 4-momentum,
$p_{0}=E^{lab}$, $p_{3}=p^{lab}=2{\pi}{\hbar}/{\lambda}^{lab}(z)$
(for the photon moving along the $z$-axis). For the cases we are discussing
the metric $g^{ij}$ can be taken in diagonal form, in particular $g^{33} =
1/g_{33}$.

From Eq. (\ref{15'})
one readily finds how ${\lambda}^{lab}(z)$ changes with height:
\beq
{\lambda}_{lab}(z)=\sqrt{g^{33}(z)/g^{33}(0)}
\sqrt{g^{00}(0)/g^{00}(z)}{\lambda}_{lab}(0) \;
\label{16'}
\eeq
On the other hand, the grating spacing in the $z$-direction, $l^{lab}(z)$,
also changes with height. This is just the standard change of scale
in the gravitational field, explained e.g. in the book by L. Landau and
E. Lifshitz, {\S} 84, \cite{nn}:
\beq
l^{lab}(z)=\sqrt{-g^{33}(z)}l^{0}
\label{17'}
\eeq
where $l^{0}$ is the ``proper spacing'' in $z$-direction,
counterpart of the proper period of the standard clock.
Thus the spacing $l^{lab}(z)$
depends on $z$ as follows
\beq
l^{lab}(z)=\sqrt{g^{33}(z)/g^{33}(0)}l^{lab}(0)
\label{18'}
\eeq
Finally, in the wavelength analogue of the Pound et al. experiment
with $z=h$ one would measure the double ratio
$(\lambda(h)/l(h))/(\lambda(0)/l(0))$. The result can be presented in the
form
\beq
{\Delta}{\lambda}^{lab}/{\lambda}^{lab}-
{\Delta}l^{lab}/l^{lab}=\sqrt{g^{00}(0)/g^{00}(h)}=gh/c^{2}
\label{19'}
\eeq
where ${\Delta}{\lambda}^{lab}/{\lambda}^{lab}=
[{\lambda}^{lab}(h)-{\lambda}^{lab}(0)]/{\lambda}^{lab}(0)$ and
analogously for ${\Delta}l^{lab}/l^{lab}$.
Notice that $g^{33}$ drops out from the result. This should be so because
there is a freedom in the choice of $z$-scale, and the observed quantities
cannot depend on this choice.
The Eq. (\ref{19'}) is analogous to the one describing
the Pound et al. experiments:
\beq
\Delta{\omega}/{\omega}-
\Delta{\epsilon}/{\epsilon}=\sqrt{g_{00}(0)/g_{00}(h)}=-gh/c^{2}
\label{20'}
\eeq
where ${\omega}$ is the frequency of the photon,
${\epsilon}/{\hbar}$ is the frequency of the clock (see Eq. (\ref{9}).
A word of explanation should be added about Eq. (\ref{20'}). In the
laboratory frame the first term in the left hand side is equal to zero,
\beq
\Delta{\omega}^{lab}/{\omega}^{lab}=0
\label{21'}
\eeq
as discussed in section 3, thus only the second term given by
Eq. (\ref{9}) contributes.

We would however like to stress
an important difference as compared to the case of frequency.
There one can independently measure the difference in rates of the
upper and lower clocks, $\Delta{\epsilon}^{lab}/{\epsilon}^{lab}$,
(analogue of the second term in the left hand side of
Eq. (\ref{19'})) and that was done in the
airplane experiments. Here the change of the scale,
${\Delta}l^{lab}/l^{lab}$, cannot be measured independently.
This important difference comes from the fact that the metric is static
while it is  $z$-dependent.

One has to realize that such a laboratory experiment with gratings cannot be
performed at the present state of the art in experimental physics (recall
the importance of M\"ossbauer effect in the experiments of Pound et al.).
However for the measurement of a large value of the redshift,
e.g. that of sodium spectral line from the sun, it is feasible. Such a
grating experiment was performed by J.W. Brault in 1962 and was described in
\S 38.5 in the monograph by
C. Misner, K. Thorne and J.A. Wheeler \cite{modern}. In this experiment the
wavelength of the emitted
light was fixed not by the lower grating, but by the atom on the sun surface.

\section{Conclusions}
The present article contains little original material; it is primarily
pedagogical. The gravitational redshift being, both theoretically and
experimentally, one of the cornerstones of General Relativity, it is  very
important that it always be taught in a simple but nevertheless correct way.
That way centers on
the universal modification of the rate of a clock exposed to a gravitational
potential. An alternative explanation in terms of a (presumed) gravitational
mass of a light pulse -- and its (presumed) potential energy -- is
incorrect and
misleading.  We exhibit  its fallacy, and schematically discuss redshift
experiments in the framework of the correct approach. We want to stress those
experiments in which an atomic clock was flown to, and kept at, high altitude
and subsequently compared with its twin that never left the ground. The
traveller
clock was found to run ahead of its earthbound twin. The blueshift of clocks
with height has thus been exhibited as an absolute phenomenon. One sees once
over again that the explanation of the gravitational redshift in terms of a
naive
``attraction of the photon by the earth" is wrong.

\section*{Acknowledgements}

We would like to thank V.V. Okorokov, who asked the question on compatibility
in
the framework of General Relativity of experiments by Pound et al., and the
airplane experiments. We also thank S.I. Blinnikov, A.D. Dolgov, A.Yu.
Morozov,
N. Straumann, K. Thorne and G. Veneziano for very interesting discussions.
We would like to express our gratitude especially to E.L. Schucking
for his help in substantially improving our bibliography and for his
insistence on a unified invariant approach  to both gravitational
and cosmological redshifts based on Killing vectors. We did however not
follow his advice wanting to focus the general reader's attention on the
fallacy of the wide-spread naive interpretation of the gravitational
redshift. Last but not least we want to thank J.A. Wheeler for his
encouragement. One of us (L.O.) would like to thank the Theoretical
Physics Division of CERN,
where part of this work was done, for their hospitality.


\begin{thebibliography}{99}
\bibitem{aaa} A. Einstein,  ``\"{U}ber  das Relativit\"{a}tsprinzip und die
aus demselben gezogenen Folgerungen," Jahrb. d. Radioaktivit\"at u.
Elektronik {\bf 4}, 411--462 (1907);
``\"{U}ber den Einfluss der Schwerkraft auf die Ausbreitung des
Lichtes," Ann. Phys. {\bf 35}, 898-908 (1911).

\bibitem{bb} A. Einstein, ``Die Grundlage der allgemeinen
Relativit\"{a}tstheorie," Ann. Phys. {\bf 49}, 769--822 (1916) \S 22;
 {\it The Meaning of Relativity} (Princeton University Press, New York,
1921), Eq.
 (106).

\bibitem{Pais} A. Pais,
{\it `Subtle is the Lord ...'}, {\it The Science and the Life
of Albert Einstein}
(Oxford University Press, Oxford, 1982), Chapter 9.

\bibitem{cc} R.V. Pound and G.A. Rebka, ``Apparent weight of photons,"
Phys. Rev. Lett. {\bf 4}, 337--341 (1960);
``Variation with temperature of the energy of recoil-free gamma rays from
solids," Phys. Rev. Lett. {\bf 4}, 274--275 (1960); ``Gravitational
red-shift in nuclear resonance," Phys. Rev. Lett. {\bf 3}, 439--441
(1959).

\bibitem{dd} R. Pound,  Uspekhi Fiz. Nauk {\bf 72}, 673--683 (1960);
``On the weight of photons," Sov. Phys. Uspekhi {\bf 3}, 875--883 (1961)
(English translation).

\bibitem{ee} R.V. Pound and J.L. Snider, ``Effect of Gravity on Gamma
Radiation," Phys. Rev. {\bf B140}, 788--803 (1965); ``Effect of gravity on
nuclear resonance," Phys. Lett. {\bf13}, 539--540 (1964).

\bibitem{ff} R. Vessot and M. Levine, ``A test of the equivalence principle
using a space-borne clock," Gen. Rel. Grav. {\bf 10}, 181--204 (1979).

\bibitem{ggg} J. Haefele and R. Keating,  ``Around the world atomic clocks:
predicted relativistic gains," Science {\bf 177}, 166--167 (1972);
``Around the world relativistic clocks: observed relativistic time gains,"
{\it ibid.}, 168--170.

\bibitem{hh} C. Alley et al., in {\it Experimental Gravitation},
Proceedings of the
Conference at Pavia (Sept. 1976), ed. B. Bertotti (Academic Press, New
York, NY, 1977).

\bibitem{jj} C. Will, {\it Theory and Experiment in Gravitational Physics}
(Cambridge University Press, Cambridge, 1981), Section 2.4;
``The confrontation between general relativity and experiment:
a 1992 upgrade," Int. J. Mod. Phys. {\bf D1}, 13--68 (1992);
preprint WUGRAV-95-5, gr-qc/9504017 (1995); \\
I. Shapiro,
Gravitazione Quanti e Relativit\`a.
Astrofisica e Cosmologia
(Giunti Barbera, Florence, 1979), Section 3; \\
N. Ashby and J. Spilker,
The Global Positioning System:  Theory and Applications {\bf 1},
(American Institute of Aeronautics and Astronautics Inc., Washington, D.C.,
1995),
Chapter 18; \\
J. Taylor, {\it Astronomical and Space Experiments to Test Relativity}, in
General Relativity and Gravitation (Cambridge University Press, Cambridge,
1987),
p. 214.

\bibitem{nn} W. Pauli, {\it Theory of Relativity} (Pergamon Press, Oxford,
1967),
Section 53.\\
C. M\"{o}ller, {\it The Theory of Relativity} (Clarendon
Press, Oxford, 1960), \S\S 92, 93.\\
L. Landau and E. Lifshitz, {\it The Classical Theory of Fields}
(Pergamon Press, Oxford, 1962), \S 89.\\
S. Weinberg, {\it Gravitation and Cosmology} (John Wiley \& Sons, New York,
NY, 1972),
 pp. 79--85;\\
P. Schneider, J. Ehlers, E.E. Falco, {\it Gravitational Lenses}
(Springer-Verlag, Berlin, 1996), pp. 93-97\\
N. Straumann, {\it General Relativity and Relativistic Astrophysics}
(Springer-Verlag, Berlin, 1984), pp. 96--99\\
I. Ciufolini and J.A. Wheeler, {\it Gravitation and Inertia}
(Princeton University Press, Princeton, 1995), pp. 97--108.

\bibitem{qq}  R.P. Feynman, {\it Lectures on Gravitation}, ed. Brian Hatfield
(Addison-Wesley Publ. Co., Reading, MA, 1995), Section 5.2.


\bibitem{Weil} H. Weyl, {\it Raum, Zeit, Materie}
(Verlag von Julius Springer, Berlin, 1923), 5te Auflage, p.322;
see also the 7th edition republished by J. Ehlers (Springer-Verlag, Berlin,
1988).

\bibitem{modern} C. Misner, K. Thorne and J. A. Wheeler, {\it Gravitation}
(Freeman and Co., San Francisco, 1973), Sections 7.2-7.5, 25.3, 25.4, 38.5;\\
R. W. Wald, {\it General Relativity} (University of Chicago Press,
Chicago and London, 1984) p.137;\\
H. Stephani, {\it General Relativity}  (Cambridge University
Press, Cambridge, 1996), second edition;\\
I. R. Kenyon, {\it General Relativity} (Oxford University Press, Oxford, 1996),
pp. 15--21.


\bibitem{19} Max Born, {\it Einstein's Theory of Relativity} (Dover
Publications, Inc., New York, 1962), Chapter VII, \S 11; \\
D. W. Sciama, {\it The Physical Foundations of General Relativity}
(Doubleday and Company, Inc., Garden City, NY, 1969), Chapter 5; \\
Ya. B. Zeldovich and I.D. Novikov, {\it Relativistic Astrophysics. Vol. 1.
Stars and Relativity} (University of Chicago Press, Chicago and London,
1971), pp.  89--93.

\bibitem{uu} {\it The New Encyclopedia Britannica} {\bf 20}, p.174 (1994),
15th Edition; \\ J. A. Wheeler, {\it A Journey into Gravity and
Spacetime} (Scientific American Library, New York, NY, 1990), pp. 48--53,
166--168; \\
 E. Taylor and J.A. Wheeler, {\it Spacetime Physics} (Freeman and Co., New
York,
1992), 3rd edition, pp. 258, 272; \\ D. Layzer, {\it Constructing the
Universe} (Scientific American Library, New York, 1984), Chapter 6, pp.
210--212.

\bibitem{yy} J. Schwinger, {\it Einstein's Legacy} (Scientific American
Library, New York, NY, 1986), pp. 141--153.

\bibitem{17a}
E.A. Desloge, ``Nonequivalence of a uniformly accelerating reference
frame and a frame at rest in a uniform gravitational field,"
Am. J. Phys. {\bf 57}, 1121--1125 (1989); ``The gravitational red shift in a
uniform field," Am. J. Phys. {\bf 58}, 856--858 (1990).

\bibitem{zz}
L. B. Okun, ``The Concept of Mass,"  Physics Today, pp. 31--36 (June 1989);
``Putting to Rest Mass Misconceptions,"  Physics Today, pp.
13,15,115,117 (May 1990);
Usp. Fiz. Nauk {\bf 158}, 511--530 (1989);
``The concept of mass (mass, energy, relativity)," Sov. Phys.
Usp. {\bf 32}, 629--638 (1989);
``Note on the meaning and terminology of Special
Relativity,"  Eur. J. Phys. {\bf 15}, 403--406 (1998).


\end{thebibliography}
\end{document}